# MRI-based Material Mass Density and Relative Stopping Power Estimation via Deep Learning for Proton Therapy

Yuan Gao[1], Chih-Wei Chang[1], Sagar Mandava[2], Raanan Marants[3], Jessica E. Scholey[4], Matthew Goette[1],

Yang Lei[1], Hui Mao[5], Jeffrey D. Bradley[1], Tian Liu[1], Jun Zhou[1],

Atchar Sudhyadhom[2*] and Xiaofeng Yang[1*]

[1]Department of Radiation Oncology and Winship Cancer Institute, Emory University, Atlanta, GA 30308
[2]GE Healthcare, Atlanta, GA 30308
[3]Department of Radiation Oncology, Harvard Medical School, Boston, MA 02115
[4]Department of Radiation Oncology, The University of California, San Francisco, CA 94143
[5]Department of Radiology and Imaging Sciences and Winship Cancer Institute, Emory University, Atlanta, GA

**Email**: xiaofeng.yang@emory.edu and atchar_sudhyadhom@dfci.harvard.edu

**Running title:** MRI Based Parametric Mapping

## Abstract

**Background:** Magnetic Resonance Imaging (MRI) is increasingly incorporated into treatment planning, because of its superior soft tissue contrast used for tumor and soft tissue delineation versus computed tomography (CT). However, MRI cannot directly provide mass density or relative stopping power (RSP) maps required for proton radiotherapy dose calculation. Several approaches have been made to overcome this limitation, including atlas-based electron density mapping and synthetic CT methods. Recently, artificial intelligence (AI) is increasingly incorporated into radiotherapy. It would be great interest of implanting AI into MRI-only based treatment planning to estimate mass density and RSP directly from MRI for proton radiotherapy.

**Purpose:** To demonstrate the feasibility of MRI-only based mass density and RSP estimation using deep learning (DL) for proton radiotherapy.

**Methods:** A DL-based framework was developed to discover underlying voxel-wise correlation between MR images and mass density and RSP. Five tissue substitute phantoms including skin, muscle, adipose, 45% hydroxyapatite (HA), and spongiosa bone were customized for MRI scanning based on material composition information from ICRP reports. Two animal tissue phantoms made of pig brain and liver were prepared for DL training. In the phantom study, two DL models were trained: one containing clinical T1 and T2 MRIs and another incorporating zero echo time (ZTE) MRIs as input. In the patient application study, two DL models were trained: one including T1 and T2 MRIs as input, and one incorporating synthetic dual-energy computed tomography (sDECT) images to provide bone tissue information. The DECT empirical model was chosen as reference to evaluate the proposed models in phantom and patient application studies.

**Results:** In the phantom study, DL model based on T1 and T2 MRI demonstrated higher accuracy mass density and RSP estimation in skin, muscle, adipose, brain, and liver with mean absolute percentage errors (MAPE) of 0.42%, 0.14%, 0.19%, 0.78% and 0.26% for mass density and 0.30%, 0.11%, 0.16%, 0.61% and 0.23% for RSP, respectively. DL model incorporating ZTE MRI improved the accuracy of mass density and RSP estimation in 45% HA and spongiosa bone with MAPE at 0.23% and 0.09% for mass density and 0.19% and 0.07% for RSP, respectively. For the patient study, DL model consisting of MRI and sDECT as input show appropriate estimation on bone tissue density and superior soft tissue contrast.

**Conclusion:** Results show feasibility of MRI-only based mass density and RSP estimation for proton therapy treatment planning using DL method.

# 1 Introduction

Proton radiotherapy has an advantage over photon radiation therapy because of the proton beam Bragg peak effect.[1,2] Protons stop in tissue immediately after depositing most energy, sparing surrounding normal tissue. In practice, proton treatment planning systems (TPSs) calculate dose deposited by a proton beam using three-dimensional (3D) maps of relative stopping power (RSP) or mass density. Currently, patient-specific maps of RSP acquired with volumetric imaging are used to translate Hounsfield number (HU) acquired using single-energy computed tomography (SECT) with appropriate conversions and calibration coefficients via stoichiometric calibration method.[3,4] The accuracy of this approach relies on the difference between chemical composition of patient tissue and tissue substitute phantoms used in calibration.[5,6] Since SECT cannot differentiate HU change as a result of different mass density or material chemical composition,[7] the error in mass density and RSP calculation can become significant. CT image noise and artifacts also contribute to the error in mass density and RSP calculation.[8] To account for these uncertainties, additional margins of 2.5%-3.5% of the proton beam range are added in clinical practice.[9]

Magnetic Resonance Imaging (MRI) is increasingly incorporated into treatment planning because of its superior soft tissue contrast used for tumor and soft tissue delineation versus CT. Importantly, MRI can reduce inter- and intra-observer contouring variability on many disease sites.[10,11] In most current workflows, target and organ at risk (OAR) contours defined from MRI are transferred to CT via image registration. This approach introduces geometrical uncertainties of 2-3 mm depending on disease site,[12,13] which are systematic errors and could lead to a geometric miss that compromises tumor control in proton radiotherapy.[14] Thus, MRI-only treatment planning studies have been performed in recent years which could increase geometric treatment accuracy by removing CT-to-MRI registration uncertainty,[15] and a reduction in patient received radiation, time, and cost.

However, unlike CT, MRI cannot provide mass density or RSP maps required for proton radiotherapy dose calculation. Several approaches have been proposed to overcome this limitation, including atlas-based electron density mapping[16] and synthetic CT-based methods.[17-19] Sudhyadhom et al. presented a method to determine material mean ionization potential from MRI.[20,21] Based on these methods for estimating material physical properties from MRI, Scholey et al. recently proposed a combined CT-MRI-based RSP estimation workflow that improved the uncertainty to within 1% for soft tissue.[22] That research inspired us to explore the feasibility of correlating mass density and RSP information with MRI only.

Interest in incorporating artificial intelligence (AI) into treatment planning has risen in recent years. Su et al. demonstrated that artificial neural networks achieved RSP uncertainty within 1% based on their phantom experiment using dual-energy CT (DECT).[23] Meanwhile, deep learning (DL) featured in hierarchical structures have been deployed to solve multi-physics inverse problems.[24,25] Numerous research efforts[26-29] have presented on the feasibility of incorporating DL into mass density or RSP estimation for proton therapy based on SECT, DECT, or cone beam CT (CBCT) images. It would be of great interest to implement DL networks into MRI-only treatment planning and develop a model to translate MRI to mass density or RSP directly. In this study, we propose a framework using DL to estimate material mass density and RSP for proton radiotherapy based on MRI.

# 2 Materials and Methods

## 2.1 Tissue substitute phantoms and data collection

Five tissue substitute phantoms and two animal tissue phantoms were included in this experiment. Phantom dimensions were 5.7x5.7x12.9 $cm^3$. Phantoms were placed into a cylindrical water container (8.41cm radius and 25.4 cm height) and secured with tape and Styrofoam for stability. The water container was filled with deionized water. The five tissue substitute phantoms were created to mimic skin, muscle, adipose, spongiosa, and 45% HA bone, based on each tissue's molecular composition provided in ICRU Report 44 and ICRP Report 23,and further modified by Scholey et al.[5,22,30] The tissue substitute phantoms were made from homogeneous mixtures of deionized water with specific ratios of porcine skin gelatin (protein substitute), porcine lard (fat substitute), hydroxyapatite (bone substitute) and a small amount of detergent/surfactant (SDS) to enhance homogeneity. Detailed material information is shown in Table 1. The two animal tissue phantoms were made from pig blood mixed with either finely minced pig brain or liver. The mass density of each animal tissue phantom was measured using a high precision scale (Practum313-1S, Sartorius Biotech, Germany) and volumetric pipettes. The chemical composition of each phantom was measured by combustion analysis at a specialized microanalytical facility. The mean excitation energy of each phantom was calculated using the Bragg additivity rule. RSP of the tissue substitute and animal tissue phantoms was measured with Varian ProBeam System (Varian Medical Systems, Palo Alto) and Zebra (IBA Dosimetry, Germany), as described by Chang et al.[28]. The proton energy was set at 150 MeV with a cyclotron current of 50 nA. The water equivalent thickness (WET) was calculated by taking the difference between the empty container and each phantom. RSP was calculated by dividing the WET with container width. The measured material information for tissue substitute and animal tissue phantoms are shown in Table 2.

**Table 1.** Mass percent compositions for each tissue substitute phantom.

| Tissue substitute phantoms | Water | Gelatin (Protein) | Lard (Fat) | Hydroxyapatite | SDS |
|---|---|---|---|---|---|
| Skin | 75.0 | 25.0 | | | |
| Muscle | 74.78 | 19.97 | 5.0 | | 0.25 |
| Adipose | | | 100 | | |
| Spongiosa | 26.61 | 11.83 | 47.43 | 12.81 | 1.32 |
| 45% HA Bone | 55.0 | | | 45.0 | |

**Table 2.** Measured mass density ($\rho_{meas}$), RSP, mean excitation energies ($I$), and elemental mass percent compositions for each tissue substitute and animal tissue phantom.

| Tissue substitute & animal tissue phantoms | $\rho_{meas}$ (g/cm$^3$) | RSP | $I$ (eV) | H | C | N | O | Na | P | S | Ca |
|---|---|---|---|---|---|---|---|---|---|---|---|
| Skin | 1.077 | 1.067 | 77.5 | 10.09 | 10.55 | 3.82 | 75.26 | | | 0.28 | |
| Muscle | 1.064 | 1.056 | 76.7 | 10.37 | 12.44 | 3.06 | 73.86 | 0.02 | | 0.25 | |
| Adipose | 0.936 | 0.979 | 61.9 | 12.37 | 77.70 | 0.16 | 9.77 | | | | |
| Spongiosa | 1.091 | 1.076 | 76.1 | 9.79 | 42.50 | 1.88 | 37.96 | 0.11 | 2.37 | 0.28 | 5.11 |
| 45% HA Bone | 1.417 | 1.344 | 108.5 | 6.24 | | | 67.48 | | 8.32 | | 17.96 |
| Brain | 1.028 | 1.014 | 77.3 | 10.96 | 6.02 | 0.73 | 82.24 | | | 0.05 | |
| Liver | 1.065 | 1.054 | 77.2 | 10.60 | 8.98 | 2.05 | 78.25 | | | 0.12 | |

The five tissue substitute phantoms were scanned using a 1.5T Siemens MAGNETOM Area MRI scanner using a T1-weighted volumetric interpolated brain (VIBE) Dixon sequence, used to generate T1-weighted Dixon VIBE predicted fat (T1-D-P-F) and T1-weighted Dixon examination VIBE predicted water (T1-D-P-W) images, and a T2-weighted short tau inversion recovery SPACE sequence for T2-STIR images. To maximize signal in bony tissues, the tissue substitute phantoms were scanned using a 3.0T GE SIGNA PET/MR using a zero echo time (ZTE) sequence[22] to compare with the T1 and T2 MRIs. The two animal tissue phantoms were also scanned using the 1.5T Siemens MAGNETOM Area MRI scanner to acquire T1 VIBE Dixon and T2-STIR sequences. T1 and T2 MRI matrices were 208x288x176 and 256x256x208, while ZTE MRI matrices were 256x256x60. T1, T2 and ZTE MRI acquisition parameters are shown in

Table 3. The obtained T1-D-P-W, T1-D-P-F, T2-STIR, ZTE MR phantom images are shown in Figure 1 and Figure 2.

**Table 3.** MRI T1, T2 and ZTE acquisition parameters.

| Sequence | Scanner | 3D voxel size (mm$^3$) | Repetition time-TR (ms) | Echo Time-TE (ms) |
| --- | --- | --- | --- | --- |
| T1 | 1.5T Siemens MAGNETOM Area | 1.247x1.247x1.2 | 7.76 | 2.39 |
| T2 | 1.5T Siemens MAGNETOM Area 1.5T | 1.016x1.016x1.1 | 3500 | 248 |
| ZTE | 3.0T GE SIGNA PET/MR | 0.8x0.8x1 | 700.7 | 0 |

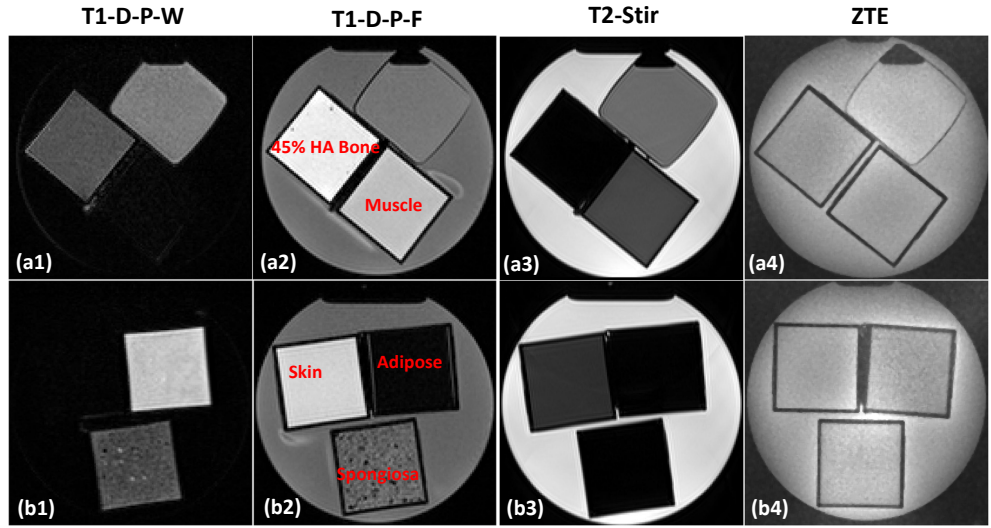

**Figure 1.** Axial images of tissue substitutes acquired from (a1)-(a4) and (b1)-(b4), T1-weighted Dixon predicated water-only (T1-D-P-W), T1-weighted Dixon predicated fat-only (T1-D-P-F), T2-weighted short tau inversion recovery (T2-stir), zero echo time sequence (ZTE).

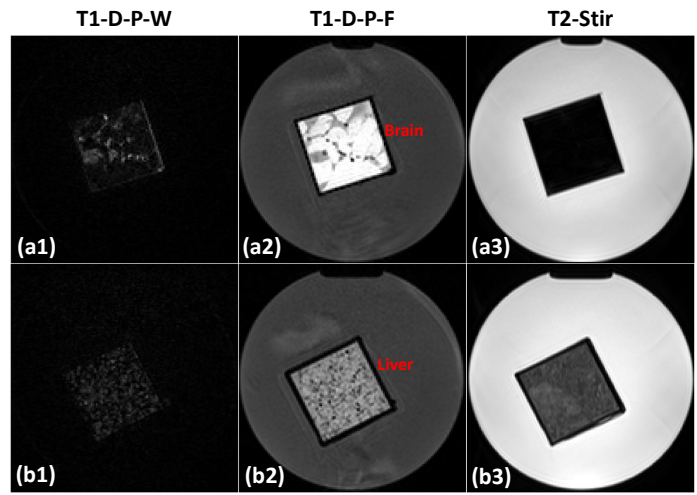

**Figure 2.** Axial images of animal tissue phantoms acquired from (a1) -(a3) and (b1) -(b3), T1-weighted Dixon predicated water-only (T1-D-P-W), T1-weighted Dixon predicated fat-only (T1-D-P-F), T2-weighted short tau inversion recovery (T2-stir).

Tissue substitute and animal tissue phantoms were also scanned with a Siemens SOMATOM Definition Edge CT scanner using a head-and-neck (HN) TwinBeam dual-energy (TBDE) protocols with a CT dose index ($CTDI_{vol, 32cm}$) of 8.6 mGy and effective milliampere-seconds ($mAs_{eff}$) of 400. In this work, DECT images acquired at 120 kVp with gold (Au) and tin (Sn) filters were referenced as DECT high-energy (HighE) and low-energy (LowE), respectively. Detailed information for DECT scans is shown in Table 4. Maps of relative electron density ($\rho_e$) and effective atomic number ($Z_{eff}$) were generated using Siemens Syngo.Via to calculate mass density and RSP as reference.

**Table 4.** DECT acquisition parameters.

| Scanner | Siemens SOMATOM Definition Edge |
|---|---|
| Collimation | 64x0.6 mm |
| Voxel size | 0.977x0.977x0.5 mm$^3$ |
| Field of view | 500 mm |
| X-ray tube voltage | 120 kVp with Au and Sn filters |

## 2.2 MRI based mass density and RSP mapping using deep learning framework

### 2.2.1 Phantom study

Both phantom and patient application frameworks were investigated for this study, as shown in Figure 3 and Figure 4. In the phantom study, two deep learning models with different inputs were investigated. The inputs of model A includes MRI T1-D-P-F, T1-D-P-W and T2-Stir images. All substitute tissue and animal tissue phantoms were trained with Model A. The inputs of model B were the same as model A, with the addition of ZTE images. We want to verify the advantage of ZTE sequence in bone material. Five tissue substitute phantoms were chosen as training targets with Model B. Mass density and RSP values estimated using Model A and B were compared in five tissue substitutes. Mass density and RSP values estimated using Model A, Model B and DECT-based reference models were compared in five tissue substitute and two animal tissue phantoms.

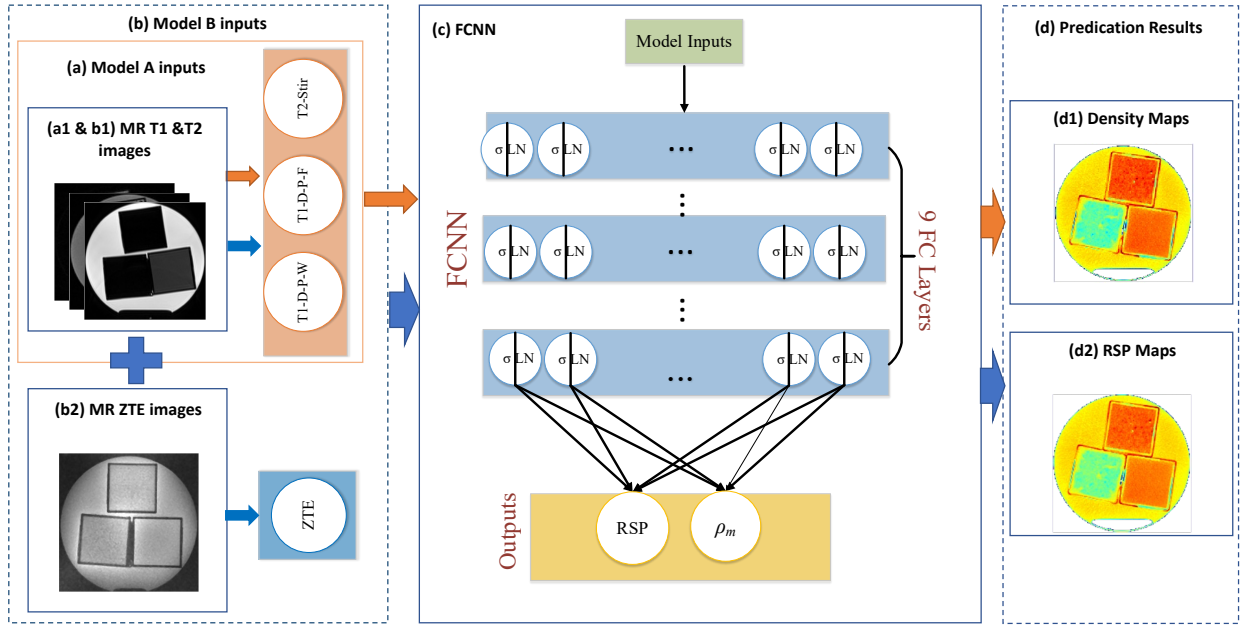

**Figure 3.** MRI-based mass density and RSP estimation with deep learning method framework for phantom study, (a) model A contains MRI T1&T2 images (volumes of interests) as input, (a1 & b1) MRI T1-D-P-W,T1-D-P-F and T2-Stir images for model A and model B, (b2) MRI ZTE images, (b) model B have MRI T1,T2 and ZTE images as inputs, (C) FCNN network structure, (d) predication results, (d1) Density maps, (d2) RSP maps. Orange arrows shows Model A inputs and outputs while blue arrows show Model B inputs and outputs.

## 2.2.2 Patient application study

Figure 4 shows the MRI-based framework for patient mass density estimation. In this part, two deep learning models (C and D) were trained with tissue phantoms' MRI and DECT images to predict mass density. In the training phase, Model C includes all 7 phantoms T1-D-P-F, T1-D-P-W and T2-Stir MR images Model C contains the same inputs with Model A, but for patient mass density map application. Model D was trained with all 7 phantoms' T1-D-P-F, T1-D-P-W and T2-Stir MRIs, DECT HighE and DECT LowE images. The patient T1-D-P-F, T1-D-P-W, T2-STIR MRIs were obtained from Velocity$^{TM}$. In the application phase, the inputs of Model C were patient T1-D-P-F, T1-D-P-W and T2-Stir MRIs. Model D includes patient T1-D-P-F, T1-D-P-W, T2-Stir and sDECT (Synthetic DECT) HighE, LowE CTs as inputs. sDECT images including sDECT HighE and LowE images were generated from MRI T1-D-P-F and T1-D-P-W images using deep learning method[31]. Which means inputs of Model D for training phase are DECT images and synthetic DECT images for application phase. Of note, there is no additional dose applied to patients in this study, since synthetic DECT is estimated from MRI. The four deep learning models based on MRI and sDECT images for phantom and patient studies are shown in Table 5.

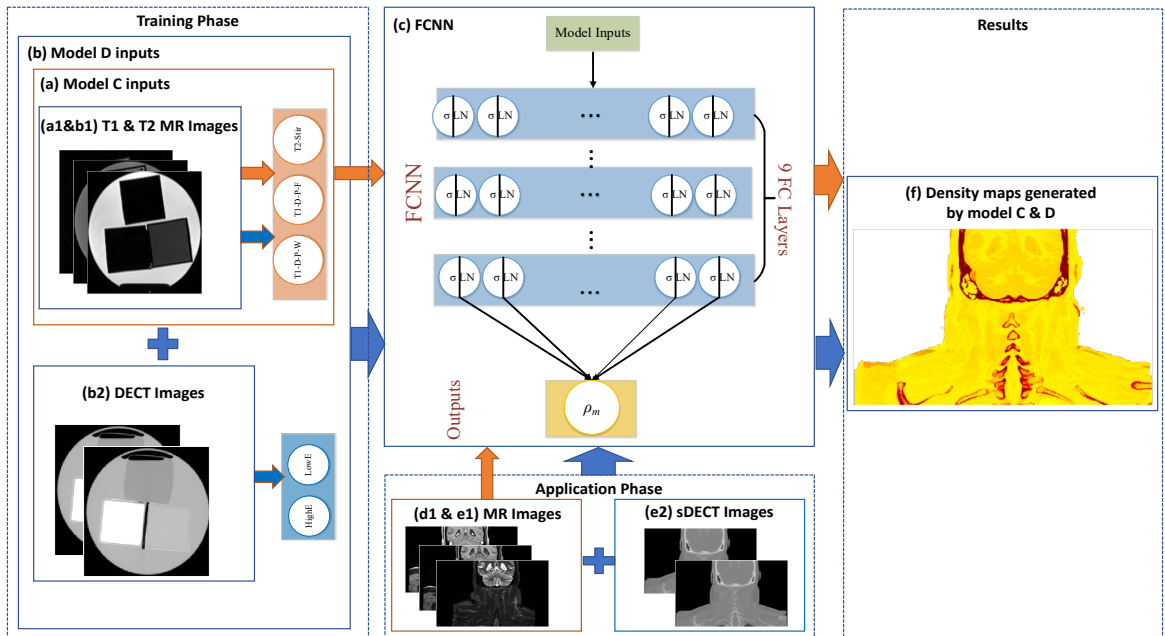

**Figure 4.** MRI-based mass density with deep learning method framework for patient application, (a) Model C has phantom MRI T1 and T2 images as input, (b) Model D has phantom MRI T1,T2 and DECT images, (a1&b1) phantom MRI T1 & T2 images for model C and model D, (b2) phantom DECT images, (c) FCNN network, (d1& e1) patient MRI T1 &T2 images, (e2) patient synthetic DECT images generated from (d1&e1), (f) patient mass density maps. Orange arrows shows Model C inputs and outputs for training phase and application phase, blue arrows show Model D inputs and outputs for training phase and application phase.

**Table 5.** Deep learning models.

| Training Set | Deep Learning Models | Inputs | Outputs |
|---|---|---|---|
| Substitute tissue & animal tissue phantoms | Model A | T1-D-P-W, T1-D-P-F, T2-STIR | Phantom mass density map, Phantom RSP map |
| Substitute tissue phantoms | Model B | T1-D-P-W, T1-D-P-F, T2-STIR, ZTE | Phantom mass density map, Phantom RSP map |
| Substitute tissue & animal tissue phantoms | Model C | T1-D-P-W, T1-D-P-F, T2-STIR | Patient mass density map |

| Substitute tissue & animal tissue phantoms | Model D | T1-D-P-W, T1-D-P-F, T2-STIR, sDECT 120 kVp Au filter (HighE), sDECT 120 kVp Sn filter (LowE), | Patient mass density map |
|---|---|---|---|

### 2.3 Deep learning model and training data preparation

Previously, we compared three deep learning models for estimating mass density and RSP based on single-energy CT (SECT) images and found that fully connected neural networks (FCNN) demonstrate better performance and lower computing cost.[29] In this study, a 1-dimension (1D) FCNN[29] was built on Pytorch[32] to perform mass density and RSP estimation based on MRI images. The details of FCNN model are shown in Figure 3 (c). Round volumes of interest (VOI) were manually contoured for 7 tissue phantoms with a radius of 15 voxels for MRI and DECT images. In the phantom study, the first half of MRI and DECT images slices were chosen as the training dataset, while the remaining half slices of images were chosen as application dataset. The number of extracted voxels in each tissue phantom is 24000. The total number of voxels in training data was 168000 for Model A and 120000 for Model B (Animal tissue phantoms were not scanned with MRI ZTE). The DL model inputs were signals from T1, T2, and ZTE MRIs at each voxel that were assigned to a 1D array with a size of 168000×3 (Model A) and 120000×4 (Model B). In the patient application study, all slices of tissue phantoms were input to the training dataset, meaning each input had 336000 voxels. The 1D array from patient images input to the model has a size of 336000×3 (Model C) and 336000×5 (Model D). The loss function used to supervise the DL models was defined as the difference between true and predicted values at each voxel.

### 2.4. Empirical model based on DECT parametric mapping

To compare DL models for estimating mass density and RSP based on MRI images, an empirical model based on DECT parametric mapping was implemented in this work. The DECT experiment set up and data collection information is shown in Table 4. The DECT parametric maps were obtained from Siemens Syngo.Via. In the phantom study, the mass density and RSP models were applied to the application dataset and compared with DL models (Model A&B) based on MR images. In the patient application study, the mass density model was applied on patient DECT images and compared with DL models (Model C&D) based on MR and sDECT images.

$$RSP = \begin{cases} \rho_e, & 0 \leq Z_{eff} < 0.5 \\ (1.1114 - 0.0148 Z_{eff})\rho_e & 0.5 \leq Z_{eff} < 8.5 \\ 0.9905 \rho_e, & 8.5 \leq Z_{eff} < 10 \\ (1.1117 - 0.0116 Z_{eff})\rho_e, & Z_{eff} \geq 10 \end{cases} \quad (2)$$

### 2.5. Evaluation method

The ground truth data was generated with the reference mass density and RSP of each phantom shown in Table 2. Absolute percentage error (APE) and mean absolute percentage error (MAPE) shown in Equation (3) and (4) were used to evaluate the performance of DL models based on MR images over each phantom, where $i$ denotes the $i^{th}$ voxel, x is the mass density or RSP at specific voxel, and N is the total voxels. In the phantom study, tables of MAPE for each tissue phantom were made to show the performance of each model.

$$APE_i = \left|\frac{x_i - x_{i,REF}}{x_{i,REF}}\right| \times 100\% \quad (3)$$

$$MAPE = \frac{1}{N}\sum_{i=1}^{N} APE_i \qquad (4)$$

## 3 Results

MAPE was used to evaluate the performance of DL models (Model A & B) based on MR images for estimating mass density and RSP for each phantom. In the patient application study, the mass maps and line profile at specific tissue were made to visualize the comparison between models (Model C & D).

### 3.1 Tissue substitute and animal tissue phantoms: mass density and RSP estimation analysis

Table 6 shows the MAPE comparisons of mass densities between MRI-based models and the DECT empirical model. DL models based on MR images (Model A & B) demonstrated better performance than the DECT empirical model over all tissue substitute and animal tissue phantoms. The DL models based on MR images also have smaller standard deviation than the DECT empirical model, except for brain. Model A has conventional T1 and T2 MR images as input, and it demonstrated the best performance in skin, muscle, adipose, brain and liver, with MAPE values of 0.42%, 0.14%, 0.19%, 0.78%, and 0.26%, respectively. In addition to the T1 and T2 MR images, inputs of Model B included ZTE MR images, and it demonstrated the best performance in spongiosa and 45% HA bone, with MAPE values of 0.23% and 0.09%, respectively. Table 7 shows the MAPE comparison of RSP between MRI-based models and the DECT empirical model. MRI-based DL models (Model A &B) demonstrated better performance than the DECT empirical model in all seven phantoms. The MRI-based DL models have smaller standard deviation than the DECT empirical model, except for brain. Model A includes conventional T1 and T2 MR images as input and demonstrated lower RSP MAPE in skin, muscle, adipose, brain and liver, at 0.30%, 0.11%, 0.16%, 0.61%, and 0.23%, respectively. The ZTE MR images were added as an extra input to Model B, which demonstrated better performance than Model A and the DECT empirical model in spongiosa and 45% HA bone with MAPE of 0.19% and 0.07%, respectively.

**Table 6.** MAPE comparison of mass densities between two MRI based FCNN models and DECT empirical model.

| Phantom | DECT Empirical | FCNN MRI Based (T1 &T2) (Model A) | FCNN MRI Based (T1, T2, ZTE) (Model B) |
|---|---|---|---|
| Skin | 0.77±0.53 | **0.42±0.21** | 0.44±0.03 |
| Muscle | 0.69±0.52 | **0.14±0.40** | 0.15±0.02 |
| Adipose | 4.61±3.38 | **0.19±0.11** | 0.26±0.07 |
| Spongiosa | 1.21±1.13 | 0.33±0.32 | **0.23±0.61** |
| 45% HA Bone | 1.47±3.00 | 0.82±2.73 | **0.09±0.70** |
| Brain | 1.73±2.50 | **0.78±3.47** | N/A |
| Liver | 0.94±1.74 | **0.26±0.27** | N/A |

**Table 7.** MAPE comparison of RSP between DECT empirical model and two MRI based FCNN models.

| Phantom | DECT Empirical | FCNN MRI Based (T1 &T2) (Model A) | FCNN MRI Based (T1, T2, ZTE) (Model B) |
|---|---|---|---|
| Skin | 0.57±0.43 | **0.30±0.16** | 0.34±0.03 |
| Muscle | 0.58±0.42 | **0.11±0.04** | 0.13±0.02 |
| Adipose | 4.34±3.28 | **0.16±0.08** | 0.24±0.05 |
| Spongiosa | 1.17±1.23 | 0.29±0.31 | **0.19±0.43** |
| 45% HA Bone | 1.32±2.76 | 0.67±2.13 | **0.07±0.55** |
| Brain | 1.43±2.40 | **0.61±2.57** | N/A |
| Liver | 0.84±1.76 | **0.23±0.19** | N/A |

## 3.2. Head-and-neck patient application study

Figure 5 illustrates the patient coronal images from T1 and T2 MRIs, DECT and sDECT HighE and LowE protocols. As shown in Figure 4 and Table 5, in the patient application phase Model C contains only T1 and T2 MRIs, as shown in Figure 5(a)-(c). Model D has sDECT images as extra input, as shown in Figure 5(f)-(g). Figure 6 (a)-(c) illustrates the mass density maps of the patient generated using different models and Figure 6(d)-(f) illustrates the mass density line profile from the red, blue, and green line shown in Figure 5(a)-(g). Figure 7(a) illustrates the mass density line profile around the skull. At voxel 25, Model C estimated a mass density of 1.04 g/cm$^3$, the DECT empirical model estimated a mass density of 1.93 g/cm$^3$, while the estimation of Model D is 1.78 g/cm$^3$. Based on ICRP 70,[33] the mass density of hydrated trabecular bone is around 1.87 g/cm$^3$. Model D and the DECT empirical model gave a close estimation for skull mass density, while Model C failed. From voxel 1 to voxel 15, the average mass density of soft tissue predicted by Model C, Model D and the DECT empirical models are 1.03 g/cm$^3$, 1.06 g/cm$^3$ and 1.05 g/cm$^3$, respectively. Based on ICRU 44,[5] typical soft tissue has a mass density range from 1.04 g/cm$^3$ to 1.07 g/cm$^3$. Predictions from the three models gave good agreement with the reference value. Figure 7(a) portrays the mass density estimation around muscle and adipose. From voxel 15 to 25, the average mass density predication by Model C, Model D and DECT empirical are 1.05 g/cm$^3$, 0.96 g/cm$^3$ and 0.88 g/cm$^3$, respectively. As shown in Figure 5(a)-(c), the tissue at voxel 15 to 25 should be adipose. The standard adipose mass density is around 0.95 g/cm$^3$, according to ICRU 44.[5] The predication of Model D is 0.01 g/cm$^3$ larger than the reference value, while Model C overpredicted the adipose mass density and DECT empirical model underestimated it. Figure 7(c) illustrates the mass density profile around the clavicle bone. In the bone marrow region of voxel 6, mass density predicted by Model C, Model D, and the DECT empirical model were 1.02 g/cm$^3$, 1.06 g/cm$^3$ and 1.32 g/cm$^3$, respectively. The DECT empirical model failed to accurately predict the bone marrow density.

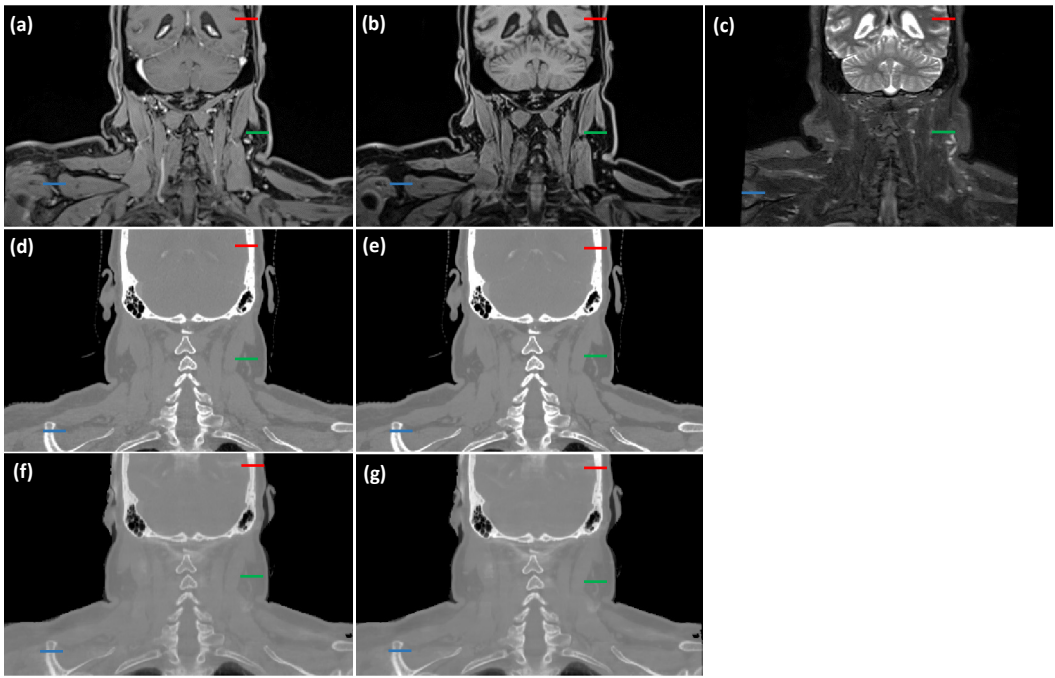

**Figure 5.** Patient coronal images from, (a)T1-weighted Dixon predicated fat-only(T1-D-P-F), (b) T1-weighted Dixon predicated water-only(T1-D-P-W), (c) T2-weighted short tau inversion recovery (T2-stir), (d) DECT high-energy, (e)DECT low-energy, (f) sDECT high-energy, (g) sDECT low-energy.

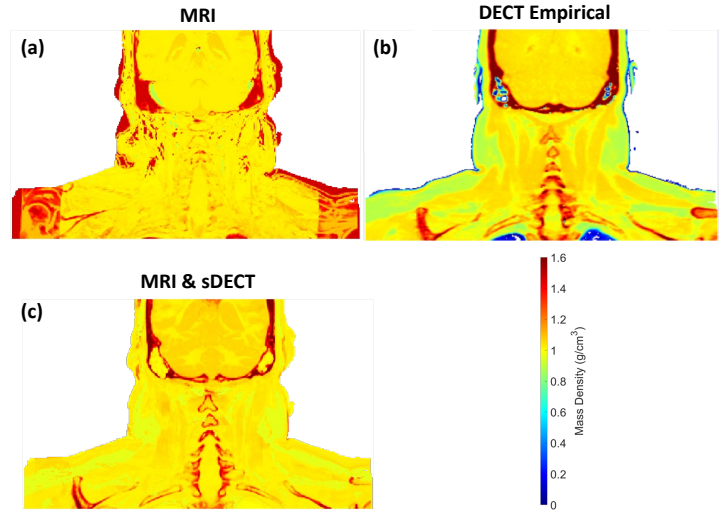

**Figure 6.** Patient mass density maps predicted from (a) T1 & T2 MRI-based (Model C) ,(b) DECT empirical model. (c) MRI and sDECT based (Model D).

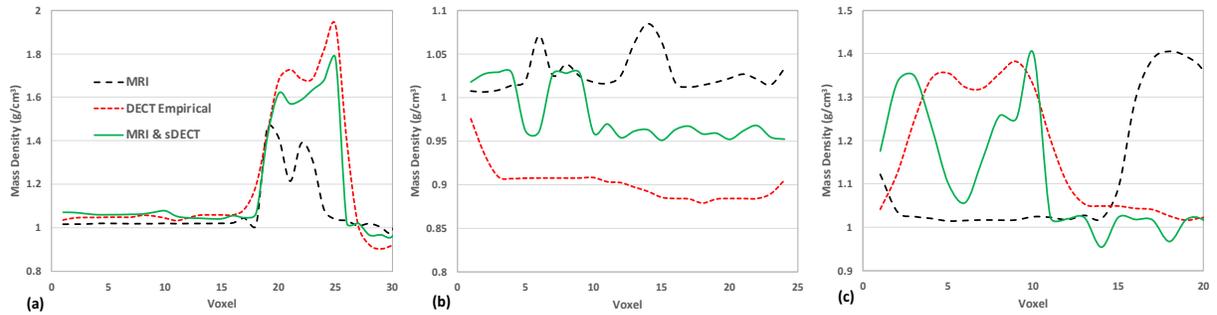

**Figure 7**. Mass density line profile (a)-(c) show the red, green, and blue line in Figure 5(a)-(g).

## 4 Discussion

The proposed MRI-based DL framework provides improved accuracy in estimating mass density and RSP of tissue substitute and animal tissue phantoms compared with traditional method. In the proposed methodology, two DL models were trained for the phantom study: Model A contains traditional T1 and T2 MRIs as input and Model B added MRI ZTE images as input. Model B demonstrates the advantage provided by ZTE MRI over conventional MRI for differentiating bone and soft tissue. In the patient application study, two models were trained with MRI and MRI-sDECT images. The MRI-sDECT model showed improvement over the conventional DECT empirical method for estimation of patient mass density.

There is a rising interest in MRI-only treatment planning in recent years. Among those methods, the deep learning based synthetic CT (sCT) method is most promising due to its superior performance.[31,34] For this approach, the workflow can be divided into two steps, the sCT images generated from MRI with DL method, and SECT stoichiometric method to build HU-mass density or HU-RSP conversion. So, the error of this approach comes from both accuracy of sCT images and the method used for calibration. Research has been performed with advanced DL models to improve the accuracy of sCT images based on MRI.[18,34,35] To overcome the limitation of SECT stoichiometric method, DECT was incorporated into treatment planning, and research has shown an improvement of mass density and RSP estimation.[36,37] Considering the advantage of DECT, sDECT generated from MRI images with DL method for treatment planning was presented.[31] In this presented methodology, we built DL models to estimate mass density and RSP based on MRI directly and created an MRI-to-mass density and MRI-to-RSP conversion, which can improve the accuracy of conventional MRI-only treatment planning (sCT, sDECT method), since this methodology bypasses the two steps that can introduce errors (error introduced by sCT). As shown in Table 1 and Table 2, Model A has improved the accuracy based on T1 and T2 MRIs, which could lead to error reduction in proton treatment planning. MRI is a multi-parametric imaging modality that not only can provide anatomical information with superior soft tissue contrast compared with CT, but also valuable functional information such as diffusion or perfusion.[15] ZTE MRI has gained more attention recently because of its superior performance for delineating bone compared with conventional T1 and T2 MRI.[38-40] In this study, Model B demonstrated better performance in estimations of mass density and RSP estimation in bone than Model A because of the additional bone signal provided by ZTE MR images.

For the patient application study, the T1 and T2 MRI-only model (model C) did not perform well in mass density estimation. We believe the main reason for this is due to the difference between patient tissue and tissue substitute phantoms, and the limitation of phantoms variability (seven tissue phantoms in total, no real bone tissue). The customized MRI-visible tissue substitute phantoms are made with water, fat or other organic materials, and their T1 and T2 relaxation timesare different than human tissue. For instance, the 45% HA phantom does not display the same contrast as cortical bone. The 45% HA phantom is dark on T2-STIR imaging, but unexpectedly bright on T1-Dxion imaging because of the microstructure difference between cortical bone and 45% HA phantom. Furthermore, the difference between human and pig tissues can also contribute the error. The other important error source comes from density and volumetric measurements, especially for the animal tissue phantoms. The MRI and sDECT model (Model D) shows appropriate mass density estimation and better performance on mass density estimation than the DECT empirical model. Of note, the proposed method offers an MRI-only framework for estimating mass density. As shown in Figure6 and Figure7, this proposed model has better performance in soft tissue estimation compared with the DECT empirical model. Bone marrow can be differentiated clearly because of the superior soft tissue information provided by MRI. For the bone tissue, the proposed model (Model D) can provide the appropriate mass density estimation because of the useful information provided by DL-based sDECT imaging. In the current state, MRI and sDECT model is the reasonable choice for patient mass density estimation.

In the future, additional customized MRI-visible phantoms will be essential for extending the training dataset and the incorporation of ZTE MRI into clinics can potentially improve the robustness of DL MRI-only model (Model C).

## 5 Conclusions

A DL framework was proposed to demonstrate the feasibility of estimating mass density and RSP based on MRI only. The DL model utilizing T1 and T2 MRIs demonstrated better performance than the DECT empirical model for estimating mass density and RSP. ZTE MRI can improve the DL model performance of estimating mass density and RSP in bone. The MRI and sDECT DL model showed appropriate patient bone tissue density estimation and provided superior information in soft tissue. The proposed framework has the potential to improve the quality of current MRI-only treatment planning for proton radiotherapy by providing more accurate mass density and RSP estimation.

**Acknowledgements**
This research is supported in part by the National Institutes of Health under Award Number R01CA215718, R56EB033332 and R01EB032680.